\documentclass[manuscript]{acmart}

\copyrightyear{2022}
\acmYear{2022} 
\setcopyright{rightsretained}
\acmConference[AIES'22]{Proceedings of the 2022 AAAI/ACM Conference
on AI, Ethics, and Society}{August 1--3, 2022}{Oxford, United Kingdom}
\acmBooktitle{Proceedings of the 2022 AAAI/ACM Conference on AI,
Ethics, and Society (AIES'22), August 1--3, 2022, Oxford, United
Kingdom}
\acmDOI{10.1145/3514094.3534188}
\acmISBN{978-1-4503-9247-1/22/08}

\begin{document}
\fancyhead{}

\title{Algorithmic Fairness and Structural Injustice: Insights from Feminist Political Philosophy}


\author{Atoosa Kasirzadeh}
\affiliation{%
  \institution{University of Edinburgh}
  \country{United Kingdom}
}
\email{atoosa.kasirzadeh@ed.ac.uk}




\begin{abstract}
Data-driven predictive algorithms are widely used to automate and guide high-stake decision making such as bail and parole recommendation, medical resource distribution, and mortgage allocation. Nevertheless, harmful outcomes biased against vulnerable groups have been reported. The growing research field known as `algorithmic fairness' aims to mitigate these harmful biases. Its primary methodology consists in proposing mathematical metrics to address the social harms resulting from an algorithm's biased outputs. The metrics are typically motivated by -- or substantively rooted in -- ideals of distributive justice, as formulated by political and legal philosophers. The perspectives of feminist political philosophers on social justice, by contrast, have been largely neglected. Some feminist philosophers have criticized the paradigm of distributive justice and have proposed corrective amendments to surmount its limitations. The present paper brings some key insights of feminist political philosophy to algorithmic fairness. The paper has three goals. First, I show that algorithmic fairness does not accommodate structural injustices in its current scope. Second, I defend the relevance of \emph{structural injustices} -- as pioneered in the contemporary philosophical literature by Iris Marion Young -- to algorithmic fairness. Third, I take some steps in developing the paradigm of `responsible algorithmic fairness' to correct for errors in the current scope and implementation of algorithmic fairness.
\end{abstract}

\ccsdesc[300]{Computer systems organization~Embedded systems}
\ccsdesc[500]{Computing methodologies~Philosophical/theoretical foundations of artificial intelligence}

\keywords{Algorithmic Bias; Algorithmic Fairness; Algorithmic Justice; Structural Injustice; Distributive Justice; Feminist Philosophy; Political Philosophy; Responsibility; Ethics of Artificial Intelligence; Ethical Machine Learning}

\maketitle

\section{Introduction}

Consequential social and institutional decisions are increasingly guided and automated by artificial intelligence, especially by resorting to data-driven and predictive machine learning algorithms. These algorithms are extensively employed to inform high-stakes decision making in a range of contexts, including the allocation of employment opportunities and the provision of medical care. Corresponding to the increased utilization of data-driven predictive machine learning in diverse social fields, the concern that the algorithms may exacerbate wrongful discrimination and biases has prompted increased scrutiny. The attributed bias potential becomes especially salient in the treatment of particular social groups on the basis of socially sensitive characteristics, like race and gender. For example, it has been demonstrated that facial recognition algorithms perform less accurately on non-white women \citep{klare2012face,buolamwini2018gender}. Moreover, recidivism algorithms are claimed to be among the most systematically biased, as they arguably accord higher risk scores to non-white individuals, for instance in the US criminal justice system \citep{julia2016machine}. Or, it has been shown that language models encode discriminatory language and harmful social stereotypes against already-vulnerable groups \cite{caliskan2017semantics,weidinger2021ethical}. The insidiousness of such biases has perpetuated wrongful moral and social harms –– for example, by ensuring that Black Americans do not receive just and fair access to scarce medical resources \citep{obermeyer2019dissecting}.

By means of redress, a recent explosion of research in the rapidly growing field of `algorithmic fairness' has burgeoned (for a comprehensive introduction to algorithmic fairness, see Barocas, Hardt, and Narayanan \citep{barocas-hardt-narayanan};  for useful reviews of algorithmic fairness tools and methods, see Corbett-Davies and Goel \citep{corbett2018measure}, Mehrabi et al. \citep{mehrabi2021survey}, Chouldechova and Roth \citep{chouldechova2020snapshot}, and Mitchel et al. \citep{mitchell2021algorithmic}). The algorithmic fairness literature primarily seeks to mathematically formulate criteria of social justice with the avowed objective of remedying or preventing wrongful harms resulting from biased algorithmic outputs. These mathematical characterizations of social justice are then used at the level of data preparation, model learning or post-processing to improve the unfair biases in machine learning systems. Since social justice is the special concern of moral, legal, and political philosophy, it is warranted for researchers on algorithmic fairness to take their bearings from philosophical accounts of social justice, reflecting the best ethical, legal, or political theory thinking of the day. To date, the primary focus of philosophical investigations of algorithmic fairness research has been rooted, in one way or another, in accounts of distributive justice. With a few exceptions outlined below, the philosophical basis of algorithmic fairness research undertaken so far has primarily aimed to render precise our intuitions about distributive justice, in particular by resolving a local resource allocation problem in order to bring about an unbiased distribution of the material or computational goods discriminated against by algorithmic outputs.

While distributive justice is significant to the completely determined desiderata of algorithmic fairness, the exclusive focus on distributive accounts of justice invidiously deals with a single, narrowly construed aspect of justice. John Rawls who is one of the most influential contemporary political theorists and philosophers of distributive justice himself acknowledges that distributive justice addresses only a single instance of the application of the concept of justice \cite[p.8]{rawls1971}. Accordingly, the narrow scope of distributive justice carries over to algorithmic fairness. The crucial point is that some problems of injustice may not have distributive solutions, either because they are not the kind of problem to which a distributive solution is adequate, or because it is unclear how to accommodate the injustice issue in terms of a distribution problem, or because it is impossible to compute one. It follows that a purely distributive justice account of algorithmic fairness runs the risk of material and social inadequacy and would become devoid of social or moral significance. This predicament raises the question of how algorithmic fairness can receive a morally and politically significant formulation. I propose to return once more to political philosophy in order to explore a constructive critique of the paradigm of distributive justice with the objective of providing the necessary emendations of algorithmic fairness. 

Below, I take some steps in developing the paradigm of `responsible algorithmic fairness' designed to correct the shortcomings in the current paradigm of algorithmic fairness. In Section 2, I briefly introduce the mathematical metrics of algorithmic fairness and argue for their conceptual commitment to distributive justice. Drawing upon the works of the prominent political theorist and philosopher Iris Marion Young, I show in Section 3 that the limited focus of algorithmic fairness on questions of distribution might not accommodate a crucial dimension of social justice, namely \emph{structural injustice}. I examine in particular the reasons why structural injustice should be included in research on algorithmic fairness, and I take some steps in characterizing how the inclusion can be implemented. My examination of structural injustice reveals that seeking one-off static solutions in terms of fairness metrics is inadequate, because narrow scope is susceptible to `ethics washing' \citep{wagner2018ethics,bietti2020ethics,kasirzadeh2021fairness}, insofar as making minor mathematical changes to algorithmic outcomes is supposed to be a putatively valuable ethical or legal solution to certain algorithmic decision-making problems. I argue that this problem requires widening the scope of algorithmic fairness to include power relations, social dynamics and actors, and structures which are among the main sources of the emergence and persistence of social injustices relevant to algorithmic systems. In Section 4, I take some steps in developing a correspondingly broad-scope conception of social justice as the basis for designing just algorithmic ecosystems, receptive to both distributional and structural sources of social injustice. The basic idea is that some notions of responsibility can help to remove structural injustices that are relevant to fair algorithmic systems. In particular, I argue that Young's account of forward-looking responsibility is to be reflected and embedded into the paradigm of algorithmic fairness. The point of responsible algorithmic fairness is to build, in one way or another, social power relations and dynamics into our accounts of fair algorithmic predictions or decisions. Section 5 presents six positive practical consequences of adopting my proposed paradigm of responsible algorithmic fairness. Finally, in Section 6 I conclude the paper.

\section{Algorithmic fairness: distributive approach}

What is fair and how should we design machine learning systems which output fair predictions or decisions? Many instances of machine learning applications that have given rise to unfairness or bias in high-stakes decision domains have motivated the systematic efforts for formalizing fairness. These efforts range from measuring statistical or demographic parity \citep{dwork2012fairness} to equal opportunity \citep{hardt2016equality} to counterfactual comparisons between members of protected and unprotected groups \citep{kusner2017counterfactual}. Let us review some of these mathematical formalizations by focusing on an example.

Few computational decision systems have given rise to as intense a discussion as COMPAS (Correctional Offender Management Profiling for Alternative Sanctions). Central to COMPAS is a recidivism risk-assessment algorithm for judging pretrial decisions concerning bail or parole. The risk-assessment algorithm predicts the likelihood that pretrial defendants will commit a violent crime in the near future or fail to appear in court. Risk-assessment decision systems similar to COMPAS are increasingly utilized in parts of the United States specifically to reduce wrongful discrimination in the criminal justice systems.\footnote{For a systematic examination of the COMPAS risk assessment tool, see Berk et al. \cite{berk2021fairness}.} 

The 2016 ProPublica analysis of the use of COMPAS in criminal justice in Broward County, Florida, claims that the algorithm's predictive results exhibit patterns of unjust racial bias \citep{julia2016machine,larson2016we}, suggesting that the performance of the COMPAS model unjustifiably varies relative to diverse racial groups. In particular, Angwin et al. \cite{julia2016machine} showed that, among the defendants who ultimately did not reoffend, Blacks were labelled `high risk' by the algorithm approximately twice as frequently as whites. By contrast, among defendants who ultimately did reoffend, whites were more than one and a half times likelier than Blacks to be labelled `low-risk' by the algorithm. Angwin et al.'s crucial observation is that the algorithm yields unjust racial bias because it manifests asymmetric predictive accuracy: it does not attribute the same false positive and false negative rates, on average, across the partition by race.

ProPublica's claim has been met with several objections. These query the specific statistical criterion for assessing the uniformity of false positive and false negative rates across the partition. The opponents defend an alternative criterion as the best metric for measuring unfair algorithmic allocations. Notably, Flores et al. \cite{flores2016false} defend the well-calibration criterion as the proper metric for measuring harmful algorithmic bias. Flores et al. \cite{flores2016false} reject the ProPublica analysis on the ground that for each possible COMPAS risk score, the percentage of blacks assigned that risk score who recidivated was approximately the same as the percentage of whites assigned the same risk score who recidivated, meaning that the algorithm was in that sense equally well-calibrated. That is, for each possible risk score, the percentage of defendants assigned a risk score class who committed a crime was statistically the same. Dieterich et al. \cite{dieterich2016compas} claim that the predictions of COMPAS are not racially biased because the predictions were equally accurate, on average, across the partition. Fundamentally, the counterproposals raise the question of which statistical criteria adequately and conclusively assess the algorithm's predictive performance (and how they are justified, not just in mathematical terms but in a substantive moral or philosophical sense). COMPAS is among several algorithms that (arguably) evince the exacerbation of unfairness and injustice through algorithmic decision making. Another recent case for `The Secret Bias Hidden in Mortgage-Approval Algorithms' revealed that, in the United States, loans to non-white applicants were 40\% to 80\% more likely to be denied relative to the white counterpart class \cite{martinez21}. In metropolitan areas especially, the disparity exceeded 250\%. For a collection of examples of unfair algorithmic biases, see Barocas, Hardt, and Narayanan \cite{barocas-hardt-narayanan}.

When viewed from a conceptual perspective, most mathematical metrics of algorithmic fairness are inherently rooted in a distributive conception of justice in that they are concerned with how to allocate the relevant computational or material goods across different groups or individuals (I discuss some exceptions in the next section). 
Broadly put, distributive justice is concerned with the institutional distribution of benefits and burdens across members of society \cite{rawls1971,fleischacker2009short,lamont2017distributive}. In research on algorithmic fairness, theories or principles of social justice are often translated into the distribution of material (such as employment opportunities) or computational (such as predictive performance) goods across the different social groups or individuals known to be affected by algorithmic outputs. 

Computer scientists do not always and explicitly provide a moral or philosophical justification of their proposed mathematical formulation of algorithmic fairness. However, across the board, whether they are explicitly defended as such or not, most mathematical criteria for algorithmic fairness can be understood as mathematical translations of some ideals or principles for distributing computational or material benefits or burdens among people. For instance, recall the statistical criterion of calibration within groups. The purpose of this criterion is to satisfy the distribution of risk scores across different groups according to the following definition: For each possible risk score, the percentage of individuals assigned that risk score who are actually positive (the individuals who go on to commit crimes) is the same for each relevant group and is equal to that risk score. In a similar vein, the equal distribution of predictive accuracy of false positive or false negative error rates across the partition by race has been the criterion of algorithmic fairness as proposed by Angwin et al. \citep{julia2016machine}.

Perhaps the most influential paradigm of distributive justice in the contemporary literature is due to John Rawls. Rawls' philosophical work has motivated an extensive program of research on algorithmic fairness. Indeed, Rawls' principles of distributive justice are so foundational to the field that he has been named `AI's favourite philosopher' \cite{Procaccia19}. Several proposed formalizations of Rawlsian theories of fair equality of opportunity are examined by Hardt et al. \cite{hardt2016equality}, Joseph et al. \cite{joseph2016fairness}, and Heidari et al. \cite{heidari2018fairness,heidari2019moral} (among others), while portions of Rawls' theory of distributive justice are given mathematical expression in the work of Joseph et al. \cite{joseph2016rawlsian} and Hoshimoto et al. \cite{hashimoto2018fairness}. More recently, Rawls' principles of distributive justice are used to offer a theory of justice for artificial intelligence \cite{gabriel2022toward}.

Several political philosophers have contributed to defending a version of the distributive justice paradigm by proposing an alternative theory of distributive justice or emending Rawls' original formulation \cite{nozick1973distributive,dworkin1981equality,dworkin1981equality2}. The relevance of their works to research on algorithmic fairness is acknowledged in the algorithmic fairness literature, most notably by Binns \cite{binns2018fairness}. While a thorough engagement with the substantial literature on the nuances of distributive justice is beyond the scope of this paper, it suffices to say that most fairness constraints and metrics are deeply rooted in the paradigm of distributive justice, broadly construed. Except for a few proposals for taking procedural fairness seriously \cite{grgic2018beyond,hoffmann2019fairness,zimmermann2021proceed}, algorithmic fairness is firmly embedded in the distributive paradigm. 

But would a fair distribution of an algorithm's predictions or decisions mitigate injustices raised by the algorithmic systems? Not always, if we truly pay attention to the sociotechnical nature of algorithms. In the next section, I defend this contention by arguing that structural injustices are relevant to our conceptions of algorithmic fairness. The importance of the distributive fairness metrics depends on the power structures and social dynamics in and through which the algorithm operates. 

\section{Algorithmic fairness and structural injustice}

Consider Mandy, a low-paid employee and single mother of two underage children. Mandy lives in a fast-growing city. As the evidence attests, a notorious Tech Giant has a vested interest in automating services in various institutions in Mandy's city. The presence of the Tech Giant contributes to driving up housing costs locally, making it unaffordable for many to live in their present homes. Greater numbers of people cannot afford rent, let alone home-ownership. Driven out of their homes, the indigent workers resort to makeshift homeless shelters under road bypasses or camp out in the parks at night, just as the Tech Giant's employees, who, bolstered by generous wages, race to buy every piece of real estate in sight.

One day, Mandy's landlord lands her with a notice of eviction, intending to renovate the premises with a view to raising the rent later. It does not take long for Mandy to realize that all landlords are pursuing similarly amoral profit-seeking rents, and she is effectively priced out of the rental market. Given her limited options, she resorts to applying for a bank loan, now allocated on the basis of algorithmic services thanks to the involvement of the Tech Giant.

Apparently, the Tech Giant has made living conditions significantly worse for Mandy and most other ordinary residents, apart from the property-owning class, whose windfall gains in real estate are astronomical. Eventually, the Tech Giant has to reckon with public pressure to mitigate this starkly unjust situation (perhaps a local grassroots movement to secure affordable housing has sprung up in Mandy's town). But how? 

One solution is to distribute a number of low-interest loans to residents. A panel of in-house hired ethicists propose that the loan allocation algorithm should comply with an equality of opportunity fairness metric. Having applied for a loan, Mandy receives the maximum amount, thanks to the distributive fairness metric embedded in the design of the decision-making algorithm. The Tech Giant's press release boasts that it complies with the ethical principles of social justice, having thus operationalized social justice into the automated decision making: ``Through our firm commitment to ethical principles, we have operationalized social justice in automated decision making.''

Let us examine this scenario more closely. First, the example clearly illustrates that algorithms are not purely mathematico-computational, do not operate in a vacuum, and cannot be analyzed in isolation. As several scholars have argued, algorithms are inherently sociotechnical entities (see, for example, Suchman \cite{suchman1987plans}, Collins \cite{collins1990artificial}, and Selbst et al. \cite{selbst2019fairness}); roughly put, algorithms are neither constructed nor adequately represented independently of the specific social contexts, institutions, power relations, and values in which they operate. On the contrary, an operating algorithm is an entity continuous with its social setting: i.e., a sociotechnical entity. The resource allocation algorithm is not only developed within an ecosystem of socio-political and ethical values, but it is also a proper component of the institutions in which it operates. We take algorithmic fairness in general to be concerned with such sociotechnical systems \cite{kasirzadeh2021ethical,dolata2021sociotechnical}. Thus, fairness of algorithms is inherently (and therefore inescapably) connected with its reception in its social context of utilization and its response to the social injustice at work.

But what can a purely local resource allocation achieve against the background of power relations and social dynamics in which the algorithm operates in? Even if, by some mathematical measure, the loan allocation algorithm is deemed `fair', it seems that the locally just decision to allocate the loan to Mandy might be fully irrelevant: it has no useful broader social significance, so long as the underlying injustice goes unaddressed. Under the conditions in which Mandy is living, it seems likely that she will accumulate a mountain of debt, instead of quickly paying off her bank loan. Overall, the initial putatively `fair' allocation will have worsened her position in the long run, despite marginally improving it for a brief time. Even if the algorithmic fairness criterion positively discriminates in favor of people like Mandy, her problem persists.

Therefore, I argue that we cannot ascertain whether `fair' distributions computed by `fair' algorithms have genuine moral force and significance except in comparison to an overall assessment of the kinds of injustices algorithms as sociotechnical systems are an important part of. In Mandy's case, the algorithm is embedded in the power relations between the Tech Giant and other individual and collective elements in that society and any assessment of fairness must be considered in light of the algorithms' capacity to address the sources, as well as outcomes, of injustice. 

Algorithms understood in terms of their sociotechnical character can act both as the exacerbating cause and the mitigating remedy of the relevant sources and patterns of social injustice. However, not all problems of social injustice have distributive solutions. While distributive matters ought to be taken seriously by algorithmic fairness research, the ethical credibility of the algorithm is moot if the algorithm itself does not address the sources of structural injustice, as opposed to merely addressing its effects. That is, genuine ethical force arises from tackling social injustice at the root, not just in softening over its effects via purely distributive (viz. outcome-based) solutions. Since distributive solutions are outcome-based, they are inherently incapable of preventing the perpetuation of long-lasting cycles of inequality. In sum, taking the sociotechnical character of algorithms seriously goes some way towards resolving the issues arising from the application of fairness metrics narrowly construed. That algorithms are sociotechnical processes entails that algorithms must be responsive to the relevant sources of social injustice. The broader inference, moreover, is that algorithmic interventions (however construed) must have a correct justification and motivation in order to be candidates for fairness or justice.

The prominent political and feminist theorist and philosopher, Iris Marion Young, popularized the contemporary concept of structural injustice, in particular as a critique of the scope of the distributive justice paradigm \cite{young2000inclusion,young2001equality,young2002lived,young2006responsibility,young2010responsibility}.\footnote{Young's conceptions of structure and structural injustice are rooted in the works of philosophers and sociologists such as Peter Blau \cite{blau1977inequality}, Pierre Bourdieu \cite{bourdieu1984distinction}, and Anthony Giddens \cite{giddens1979central}. See Young \cite{young2006responsibility} for more details.} Young's argument begins with the observation that contemporary liberal political philosophy has been narrowly and merely focused on distributive justice. But, she argues, beyond concerns about distribution, unjust and wrongful harms against some people, but benefits for others, are also the result of structures and processes. According to Young, we might not be able to resolve these structural harms simply by altering the distribution of burdens and benefits across the members of a society. In these cases, the distributive approach alone might become only cosmetic. Rather, cases of structural injustice demand that we overcome, revise, or reform the problematic structures.

In Mandy's type of scenario, the structure of injustice is not necessarily identifiable by isolating individual responsible agents. Rather, the injustice arises as the sum of (arguably) non-blameworthy actions of multiple agents which can only be changed through collective action. Let us call such injustices `structural', as Iris Marion Young proposes \cite{young2006responsibility}. What are these structures and when do structural injustices exist?

The structure, according to Young \cite[p.111]{young2006responsibility}, is defined as follows:

\begin{quote}
As I understand the concept, the confluence of institutional rules and interactive routines, mobilization of resources, as well as physical structures such as buildings and roads. These constitute the historical givens in relation to which individuals act, and which are relatively stable over time. Social structures serve as background conditions for individual actions by presenting actors with options; they provide ``channels'' that both enable action and constrain it.
\end{quote}

Moreover, in a widely cited passage, Young \cite[p. 52]{young2011justice} defines structural injustice as follows:

\begin{quote}
Structural injustice, then, exists when social processes put large groups of persons under systematic threat of domination or deprivation of the means to develop and exercise their capacities, at the same time that these processes enable others to dominate or to have a wide range of opportunities for developing and exercising capacities available to them. [...] Structural injustice occurs as a consequence of many individuals and institutions acting to pursue their particular goals and interests, for the most part within the limits of accepted rules and norms.
\end{quote}

That is to say, structural injustice results from the operation of multiple social processes; in Mandy's case, this includes the unregulated expansion of the economic and social power of a tech giant coupled with laissez-faire policies and unregulated profit-oriented capitalism. The aggregate of factors is ultimately responsible for Mandy's situation. The situation is familiar enough, as it places individuals and groups under the systematic threat of domination or deprivation, while enabling others to dominate and exploit them.  

To reiterate, Mandy's scenario illustrates an example of structural injustice caused by a supremely powerful individual company as well as other actors in positions of power (the mayor, the city council, etc.). Note that structural injustice renders individuals (and certain social groups) vulnerable to domination or oppression, even if there is no individual to blame. In Mandy's case, conceptualizing problems of justice as `goods' that must be distributed partially obscures the fact that many actions which are the roots and the reasons for the persistence of the relevant injustice result from dynamic processes and relations.

If Tech Giant is actively contributing to the unjust social conditions besetting Mandy and others like her, the algorithmic distributive fairness of the loan allocation itself is largely beside the point, since it is the source of the injustice (and its perpetuation) that matters. To put it differently, although some fairness criteria, e.g., those proposed to address statistical error distributions, can be justified locally in terms of ensuring fairer outcomes, the social significance of the fair algorithm is negligible inasmuch as it fails to address (or even recognize) the cause of the injustice in the Tech Giant's activities. To guarantee significance, the dynamic and compounding causal elements of injustice must be included into the assessment of algorithmic fairness. The key cause of social injustice is the negative change in living conditions brought about by the Tech Giant. 

According to Young \cite{young2010responsibility}, structural injustice is a moral wrong distinct from the wrongful action of an individual agent or the repressive policies of a state. In addition, structural injustice contrasts with at least two other forms of harms or wrongs, namely that which comes about through an individual action, and that which is attributable to specific actions and policies of states or other powerful institutions. Thus, structural injustice occurs as an aggregate consequence of individuals and institutions acting to pursue particular goals and interests, largely within the limits of accepted rules and norms.

To be sure, the limits of distributive fairness metrics have not been entirely neglected by the literature on algorithmic fairness. Grgić-Hlača et al. \cite{grgic2018beyond} proposes ways to include procedural fairness, i.e., the fairness of the decision making, into the research on algorithmic fairness. Hoffmann \cite{hoffmann2019fairness} draws heavily on legal studies in discussing the limits of the goods included in the prevailing paradigm of distributive justice in the literature. Zimmermann and Lee-Stronach \cite{zimmermann2021proceed} draw attention to the relevance of procedural justice to algorithmic systems and argue that reliance on algorithmic systems is procedurally unjust in contexts involving background conditions of structural injustice. The arguments presented in this paper are novel in that Grgić-Hlača et al. \cite{grgic2018beyond} does not engage with the literature on structural injustice. Moreover, while Hoffmann \cite{hoffmann2019fairness} and Zimmermann and Lee-Stronach \cite{zimmermann2021proceed} briefly quote Young's work on structural injustice, they do not explicitly engage with her account for correcting the limitations of fairness metrics. In this paper, I complement these earlier works by arguing why and how structural injustice is to be conceptually included in algorithmic fairness research.

\section{Towards responsible algorithmic fairness}

Granting the broad lines of argument defended so far, the following question arises: How should sociotechnical algorithmic systems deal with structural injustice? How should the designers of the algorithm, e.g., the Tech Giant employees, and the Tech Giant itself act in light of the existence of structural injustice?

In correcting the harms of structural injustices, Young suggests that the key is to develop an appropriate account of responsibility, which she calls a \emph{social connection model} \cite{young2006responsibility,young2011justice}. The social connection model of responsibility is forward-looking; it says what can be done when we go forward.\footnote{According to Young, it is undesirable to rely on a purely legalistic liability model to restrict the range to cases in which an agent causing the harm may be identified. Young suggests to shift from mere attribution of blame and guilt to an action model. How these two models relate to each other is itself a subject of ongoing discussion. See, for example, \cite{goodin2021responsibility}.} 

According to Young, agents who participate in a structure of concerted action (e.g., a cooperative structure) are responsible for them in the sense that they are part of the process that causes structural injustice further down the line. In the case of Mandy, the responsible actors include the owners and designers of the algorithms, the Tech Giant as institution, its employees, and similar agents. While these actors are not necessarily responsible in the sense of having directed the process or intended the outcome, they nevertheless bear responsibility for the resulting structural injustice, because they contribute by their actions to the processes that produce unjust outcomes.

Young suggests four modes of reasoning for engaging with the social connection model of responsibility for structural injustice. According to her, responsibility for justice encourages us to think about how we can best take responsibility for reducing injustice by reflecting on these four parameters: different options of power, privilege, interest, and collective ability. 

First, take the concept of power. Different levels of causal influence and capacities are relevant and contribute to changing processes. For example, a few key players may have both a greater capacity to make changes themselves and to influence the process. A few select actors are responsible for the choice of fairness metric; part of their responsibility ought to be to assess whether the selected metric contributes to structural injustice. Equally, it is the responsibility of corporations with both vast influence and capacity to mitigate structural injustices arising operationally, in this scenario, from machine learning algorithms or other technical systems. 

Second, consider the concept of privilege. Privilege also influences one's degree of responsibility: the more privileged have a greater responsibility to address injustice. By virtue of their social position, the privileged may, like the powerful, exert an influence on the adoption of just fairness metrics in algorithms. They hold an additional responsibility to devote some of the resources—material or otherwise—towards the cause of justice in this field.

Third, take the notion of an interest vested in the structure. Those who have a vital interest in changing oppressive structures have direct responsibilities in the order of mitigation. To the extent that the powerful have a vested interest in addressing injustice, they too must examine the broader fairness consequences of algorithmic systems. 

Fourth, consider collective ability. In some cases, collective organization capacities and resources exist to some extent. It is a reasonable practical policy to draw upon these (if they exist at all). In some instances, student associations, faith-based organizations, labor unions, or stockholder groups already exercise some degree of power in being able to coordinate members to take certain actions. Young proposes to harness organizational resources where doing so would prove effective. 

The degrees and kind of forward-looking responsibility for changing structural processes to reduce injustice vary according to the four parameters sketched above. On Young's account, No one who participates in processes that produce structural injustice is exempt from responsibility to join with others to change those structures. Some are less inclined to do so, however, because their positions give them more interest in preserving than in changing them. Others stand in positions of relative weakness in the structures. The former must usually be pressured to take steps aimed at changing the effects of their actions, and the political responsibility of the latter often amounts to little else than organizing to criticize and pressure more powerful actors \cite{young2010responsibility}. As she continues, it requires organization, the will to cooperate on the part of many diverse actors, significant knowledge of how the actions of individuals and the rules and purposes of institutions conspire to produce injustice, and the ability to foresee the likely consequences of proposed remedies. One or more of these conditions is often absent. But something else often stands in the way of trying to bring about these conditions, namely the attempt by participants in the process to deny that they have a responsibility to try to remedy injustice. Future work is required to produce more awareness about such strategies of avoidance in talking to one another about responsibility and collective action, and in holding one another accountable.

Yet, unlike retrospective responsibility, forward-looking responsibility is not a matter of having caused an existing (morally problematic) state of affairs. Instead, Young's account renders responsibility a matter of being morally charged with—that is, responsible for—bringing about a state of affairs that we collectively regard as an improvement. Forward-looking responsibility is not completely removed from considerations of blame -- we sometimes appropriately blame those who fail to take their responsibilities seriously—but blame is not morally salient. Rather, prospective responsibility is morally salient because we think that such responsibility may, if taken seriously by those who are held responsible, help to bring about an improved worldly state of affairs.

Given the power and privilege of many institutions who design artificial intelligence systems, it is necessary to address the four parameters of the problem as proposed by Young. That is to say, the designers and social institutions owe a four-way responsibility to address structural injustice. As illustrated and argued above, AI governance requires responsibility in this comprehensive sense.

\section{Practical consequences}

Next, I examine the practical consequences for algorithmic fairness. 

First, the motivations, justifications, and significance of operationalizing relevant distributional and structural elements of social injustice into algorithmic ecosystems are to be evaluated more carefully and systematically. Algorithms are to be understood as sociotechnical entities in a rich sense of the term. The range of issues that can legitimately be discussed and evaluated in relation to the sociotechnical character of algorithmic fairness must be comprehensively expanded, while finessing the representation of the relevant concepts of social (in)justice. Specifically, algorithmic fairness is to be evaluated in a broader context of algorithmic justice, encompassing both the distributive and structural elements of injustices. This in part requires broadening the understanding of justice for sociotechnical systems by regarding algorithms not as mere distributors of computational or material goods, but as participants, doers, and actors embedded in complex social relations and dynamics.

Second, algorithmic outputs by informing or automating various decisions affect lives and have significant social impacts. Hence, they are intrinsically bound up with structures of social power. Discussions of algorithmic fairness tend to often overlook this power aspect of the operative context. I, however, showed that social justice concerns and formal notions of algorithmic fairness must be jointly analyzed, and that the promises and limits of the formal properties of fair predictions or decisions must be robustly related to the context of social power. A socially situated conception of algorithmic justice and fairness allows us to better trace the interdependencies between power, algorithmic reason, algorithmic rationality, and authority, and to consider how power relations affect the mode as well as the outcome of algorithmic applications. 

Third, the areas of intervening to secure social justice should not be merely local and technical, merely at the level of data preparation, model learning or post-processing. To know whether an algorithmic output is fair it should be evaluated in trade-off with other interconnected solutions to social justice. This discussion of how to approach this trade-off problem should find central stage and be taken seriously while specifying agendas for algorithmic fairness research program. To this end, future interdisciplinary work is needed to give more precision to the notion of structural injustice and to find a meaningful trade-off between its demands in the context of algorithmic predictions or decisions.

Fourth, algorithmic fairness must be treated in relation to responsibility. How should we design fair algorithms responsibly? To this end, I took some steps forward in applying Young's fourfold account of forward-looking responsibility. The interrelation between the topics of moral and political responsibility and algorithmic fairness which have gone largely neglected in the literature are to be articulated together. Responsible algorithmic fairness renders the notions of structural and distributive justice as complements to each other, not competitors.

Fifth, the proposed sense of responsible fairness brings the discussion to the notion of the responsible citizen or the responsible corporations and companies. As individual agents, overspecialization threatens to obscure the true extent of personal responsibility. A computer scientist's role, for example, is not only to design a fair algorithm by measuring some local aspects of fairness via mathematical and computational tools. In her role as a citizen, she is also responsible for working against structural injustices in her capacity. This conception requires collaboration between different ethical, social, and political groups. The broader understanding of justice brings considerations of power to the fore. These, rather than the purely formal properties of metrics, are the fundamental issues of relevance to algorithmic fairness to address. 
	
Finally, algorithms are powerful tools with a range of uses. In application, they promote good or bad outcomes, depending upon implementation and context. I think there is the potential that algorithmic systems can be used to repair some problematic structures and to generate better ones. It is essentially the perspective of structural justice that will allow us to determine the applications of algorithmic systems with genuine moral force.

\section{Conclusion}

Machine learning algorithms are widely used to guide the distribution of goods. Within the arena of machine learning, algorithmic fairness is a burgeoning field of research that focuses on the design of mathematical metrics for redressing harms caused by biased distributional effects of societal algorithmic systems. These metrics are often motivated by the twin political ideals of social justice and fairness. Against this background, the current paper drew on feminist political theory in arguing that a primarily distributive approach to the allocation of material and computational goods is too narrow to adequately address or accommodate concerns about social injustice vis-à-vis algorithmic fairness. Algorithmic fairness is, therefore, morally salient and in need of philosophical attention. The paper argued that algorithmic ecosystems are socio-technical entities, and so must be receptive to different sources of social injustice. The metrics of algorithmic fairness are concerned with matters of distributional justice. However, not all sources of social injustice are distributional; some are structural. After taking some steps in unpacking what responsible algorithmic fairness is, I argued for six positive corollaries of its adoption as the conceptual basis for research into the infrastructural fairness of algorithmic ecosystems and their direct effects.

This paper aimed to connect some dimensions of the philosophical works of Iris Marion Young to algorithmic fairness. I used a Youngian conception of forward-looking responsibility to sketch a way to improve the limits of algorithmic fairness. Other accounts of forward-looking responsibility have been also discussed and developed in the literature; see for example, Miller \cite{miller2008national} French and Wettstein \cite{french2014forward}. One next step is to make a more comprehensive engagement with the literature on forward-looking responsibility in the context of algorithmic justice. In this paper, I examined a particular approach for bridging feminist philosophy to the literature on algorithmic fairness. A complimentary exploration of the works of other feminist philosophers for improving the current conceptions of algorithmic fairness is an important task to explore elsewhere.

\section*{ACKNOWLEDGEMENTS}

I am grateful to Philip Pettit, Mario Günther, Iason Gabriel, two anonymous reviewers, audiences of the Philosophy Discussion Club at Cornell University, participants in the seminar on Human Rights and Global Politics at Scuola Superiore Sant'Anna Pisa, and audiences at the workshop on Responsibility \& AI at the University of Vienna for valuable discussions and suggestions.

\bibliographystyle{ACM-Reference-Format}
\bibliography{sample-base}

\end{document}